\documentstyle[preprint,tighten,aps,epsfig]{revtex}
\def\bra#1{{\langle#1\vert}}
\def\ket#1{{\vert#1\rangle}}

\def\sst#1{{\scriptscriptstyle #1}}

\def\notv{{\not\! v}}

\def\mn{{m_\sst{N}}}

\def\mws{{M^2_\sst{W}}}

\def\sstw{{\sin^2\theta_\sst{W}}}

\def\sst#1{{\scriptscriptstyle #1}}
\def\gpnn{{g_{\sst{NN}\pi}}}

\def\mpi{{m_\pi}}

\def\qw0{{Q_\sst{W}^0}}

\def\lamchi{{\Lambda_\chi}}
\def\lamchis{{\Lambda_\chi^2}}
\def\lamchic{{\Lambda_\chi^3}}
\def\hpi{{h_\pi}}
\def\hweak{{{\cal H}_\sst{W}^\sst{PV}}}
\def\gpi{{g_\pi}}
\def\hpibare{{h_\pi^\sst{BARE}}}
\def\hpieff{{h_\pi^\sst{EFF}}}

\begin{document}

\begin{titlepage}

\begin{center}

{\large{\bf Chiral Symmetry and the Parity-Violating $NN\pi$ Yukawa Coupling}}

\vspace{1.2cm}

Shi-Lin Zhu$^a$, S.J. Puglia$^a$, B.R. Holstein$^c$, M. J.
Ramsey-Musolf$^{a,b}$

\vspace{0.8cm}

$^a$ Department of Physics, University of Connecticut,
Storrs, CT 06269 USA\\
$^b$ Theory Group, Thomas Jefferson National Accelerator Facility, Newport
News,
VA 23606 USA\\
$^c$ Department of Physics and Astronomy, University of Massachusetts,
Amherst, MA 01003 USA

\end{center}

\vspace{1.0cm}

\begin{abstract}
We construct the complete $SU(2)$ parity-violating (PV) $\pi, N, \Delta$
interaction Lagrangian with one derivative, and calculate the chiral
corrections to
the PV Yukawa $NN\pi$ coupling constant $h_\pi$ through
${\cal O}(1/\Lambda_\chi^3)$ in the leading order of heavy baryon expansion.
We discuss the
relationship between the renormalized $\hpi$, the measured
value of $\hpi$, and the corresponding quantity calculated microscopically
from the
Standard Model four-quark PV interaction.

\vskip 0.5 true cm
PACS Indices: 21.30.+y, 13.40.Ks, 13.88.+e, 11.30.Er

\end{abstract}

\vspace{2cm}
\vfill
\end{titlepage}

\pagenumbering{arabic}
\section{Introduction}
\label{sec1}

The parity-violating (PV) nucleon-nucleon interaction has been a subject of
interest
in nuclear and particle physics for some time. To date, PV observables
generated by this
interaction remain the only experimental windows on the $\Delta S=0$,
nonleptonic weak
interaction. Since the 1970's, the PV NN interaction has been studied in a
variety of
processes, including ${\vec p}-p$ and ${\vec p}-$nucleus scattering,
$\gamma$-decays of
light nuclei, the scattering of epithermal neutrons from heavy nuclei, and
atomic PV
(for a review, see Refs. \cite{hh95,haxton}. The on-going interest in the
subject has spawned new PV
experiments in few-body systems, including high-energy ${\vec p}-p$
scattering at COSY,
$\vec{n}+p\to d+\gamma$ at LANSCE\cite{sn}, $\gamma+d\rightarrow
n+p$ at JLab\cite{jl}, and the rotation of polarized neutrons in
helium at NIST.

The theoretical analysis of these PV observables is complicated by the
short range of the low-energy
weak interaction. The Compton wavelength of the weak gauge bosons ($\sim
0.002$ fm) implies that
direct $W^\pm$ and $Z$ exchange between nucleons is highly suppressed by
the short-range repulsive
core of the strong NN interaction. In the conventional framework, longer
range PV effects arise
from the exchange of light mesons between nucleons. One requires the
exchange of the $\pi$, $\rho$ and
$\omega$ in order to saturate the seven spin-isospin channels associated
with the quantum numbers of the
underlying four-quark strangeness-conserving PV interaction, $\hweak(\Delta
S=0)$ (henceforth, the
$\Delta S=0$ will be understood). These exchanges are parameterized by PV
meson-nucleon couplings,
$h_\sst{M}$, whose values may be extracted from experiment. At present,
there appear to be discrepancies
between the values extracted from different experiments. In particular, the
values of the isovector
$\pi NN$ coupling, $\hpi$, and the isoscalar $\rho NN$ coupling, $h_\rho^0$
-- as extracted from
$\vec p-p$ scattering and the $\gamma$-decay of $^{18}$F, do not appear to
agree with the corresponding
values implied by the anapole moment of $^{133}$Cs as measured in atomic PV
\cite{apv}.

The origin of this discrepancy is not understood. One possibility is that
the use of $\rho$ and
$\omega$ exchange to describe the short-range part of the PV NN interaction
is inadequate. An alternate
approach, using effective field theory (EFT), involves an expansion of the
short-range PV NN interaction
in a series of four-nucleon contact interactions whose coefficients are
{\em a priori} unknown but in
principle could be determined from experiment. The use of $\rho$ and
$\omega$ exchange amounts to
adoption of a model -- rather than the use of experiment -- to determine
the coefficients of the
higher-derivative operators in this expansion. Whether or not the
application of EFT to nuclear PV
can yield a more self-consistent set of PV low-energy constants than the
meson-exchange approach remains
to be seen. A comprehensive analysis of nuclear PV observables using EFT
has yet to be performed.

The least ambiguous element -- shared by both approaches -- involves the
long-range $\pi$-exchange
interaction. At leading order in the derivative expansion, the PV $\pi NN$
interaction is a purely
isovector, Yukawa interaction. The strength of this interaction is
characterized by the same constant --
$\hpi$ -- in both the EFT and meson-exchange approaches. At the level of
the Standard Model (SM),
$\hpi$ is particularly sensitive to the neutral current component of
$\hweak$. In this respect, the
result of $^{18}$F PV $\gamma$-decay measurement is puzzling:
\begin{equation}
\label{eq:flourine}
\hpi = (0.73\pm 2.3) \gpi\ \ \ ,
\end{equation}
where $g_\pi = 3.8\times 10^{-8}$ gives the scale of the $h_\sst{M}$ in the
absence of neutral currents\cite{ta}.
This result is especially significant, since the relevant two-body nuclear
parity-mixing matrix element can
be obtained by isospin symmetry from the $\beta$-decay of $^{18}$Ne
\cite{haxton}. The result
in Eq. (\ref{eq:flourine}) is, thus, relatively insensitive to the nuclear
model.
Theoretical calculations of $\hpi$ starting from $\hweak$ have been
performed using SU(6)$_w$ symmetry
and the quark model \cite{ddh,fcdh}, the Skyrme model \cite{t1}, and QCD
sum rules \cite{t2}. As a
benchmark for comparison with experiment, we refer the SU(6)$_w$/quark
model analysis of Refs.
\cite{ddh,fcdh}---hereafter referred to as DDH,FCDH
\footnote{Note that although the DDH analysis used the symmetry
group SU(6)$_w$ in order to connect weak vector meson and pion couplings
the predictions relating pion couplings alone to hyperon decay data rely
only SU(3).}.  These authors quote a \lq\lq best value"
and \lq\lq
reasonable range" for the
$h_\sst{M}$:
\begin{eqnarray}
\label{eq:fcdhbest}
\hpi({\hbox{best}}) & = & 7\gpi \\
\label{eq:fcdhrange}
\hpi({\hbox{range}}) & : & (0\to 30)\gpi \ \ \ .
\end{eqnarray}
where here the \lq\lq best value" is more aptly described as an
educated guess, while the \lq\lq reasonable range" indicates a set of
numbers such that
theory would be very hard-pressed to explain were the experimental
value not found to be within this band.  Neverthess, the difference
between the
\lq\lq best value" of Eq. (\ref{eq:fcdhbest})
and the $^{18}$F result would appear to call for an explanation, and
in the following note we comment on a possible source of the
discrepancy.

In general, the problem of relating the fundamental weak quark-quark
interaction to the low-energy
constants which parameterize hadronic matrix elements of that interaction
is non-trivial. In the
framework of EFT, one may define these constants at tree-level in the
hadronic effective theory. The
quantities extracted from experiment in the conventional analysis, however,
are not the tree-level
parameters, but rather renormalized couplings. Denoting the latter as
$\hpieff$, one has
\begin{equation}
\label{eq:hpieff}
\hpieff = Z_N \sqrt{Z_\pi} \hpibare + \Delta\hpi \ \ \ ,
\end{equation}
where $\hpibare$ is the coefficient of the leading-order, PV Yukawa
interaction in
the effective theory, $\sqrt{Z_N}$ and $\sqrt{Z_\pi}$ denotes chiral loop
renormalizations of the
nucleon and pion wavefunctions, respectively, and $\Delta\hpi$ denotes
contributions from chiral loops
and higher-dimension operators to the Yukawa interactions (only the finite
parts of these couplings are
implied; loop divergences are cancelled by the corresponding pole terms in
$\hpibare$ and the
$Z_{N,\pi}$). At leading order in
$1/\lamchi$, one has $Z_{N,\pi}=1$, $\Delta\hpi = 0$, and $\hpieff =
\hpibare$. The renormalized coupling
appears as the coefficient in the one-pion-exchange (OPE) PV NN potential
\begin{equation}
\label{eq:opeint}
{\hat H}_\sst{PV}^\sst{OPE} =
i{\gpnn\hpieff\over\sqrt{2}}\left({\tau_1\times\tau_2\over 2}\right)_z
({\vec\sigma}_1+{\vec\sigma}_2)\cdot\left[{{\vec p}_1-{\vec p}_2\over
2\mn}, f_\pi(r)\right] \ \ \ ,
\end{equation}
where $\gpnn$ is the strong $\pi NN$ coupling and $f_\pi(r)= \exp(-\mpi
r)/4\pi r$. Neglecting the
effects of three-body PV forces and $2\pi$-exchange interactions, it is
$\hpieff$ to which the result
in Eq. (\ref{eq:flourine}) corresponds.

The relationship between $\hpieff$ and the coupling obtained by computing
$\bra{N\pi}\hweak\ket{N}$
in a microscopic model is not immediately transparent. In what follows, we
make several observations
about this relationship. We first show that $Z_N\sqrt{Z_\pi}$ and
$\Delta\hpi$ are substantial, so
that $\hpieff$ differs significantly from $\hpibare$. To that end, we
compute all of the chiral
corrections to the PV Yukawa interaction through ${\cal O}(1/\lamchic)$,
where $\lamchi=4\pi F_\pi$.
We work to leading order in $1/\mn$ in heavy baryon chiral perturbation
theory (HBChPT).
Of particular significance is the dependence of $\Delta\hpi$ on other
low-energy constants
parameterizing PV $2\pi$ production and the PV $N\to\pi\pi\Delta$ transition.
We subsequently reexamine the
SU(6)$_w$/quark model calculation of Refs. \cite{ddh,fcdh} and argue that
most -- if not all -- of the
chiral loop effects which renormalize $\hpi$ are not included in the
microscopic calculation. Thus,
the relationship between $\hpieff$ and microscopic calculations remains
ambiguous at best. This
ambiguity is unlikely to be resolved until an unquenched lattice QCD
calculation of $\hpi$ using light
quarks becomes tenable. In the meantime, one should not necessarily
view a discrepancy between the experimental
value of $\hpieff$ and microscopic model calculations as disturbing.

Our discussion of these observations is organized as follows. In Section 2
we summarize our conventions
and notation, including the PV chiral Lagrangians relevant to $\hpi$
renormalization. Section 3 gives a
discussion of the loop calculations. In Section 4 we comment on the scale
of the loop corrections and
provide simple estimates of some of the new PV low-energy constants
appearing in the analysis. Section
5 gives our discussion of the relationship between $\hpieff$ and the
calculation of Refs.
\cite{ddh,fcdh}. Section 6 summarizes our conclusions. Some technical
details are relegated to the
Appendices.

\section{Notations and Conventions}
\label{sec2}

We follow standard HBChPT conventions \cite{j1,ijmpe} and introduce
\begin{equation}
\Sigma =\xi^2\  ,\ \  \xi =\exp({i\pi\over F_\pi})\  ,\ \  \pi ={1\over 2}
\pi^a \tau^a
\end{equation}
with $F_\pi =92.4$ MeV being the pion decay constant.
The chiral vector and axial vector currents are given by
\begin{eqnarray}\nonumber
{\cal D}_\mu & = & D_\mu +V_\mu \\
A_\mu &=& -{i\over 2}(\xi D_\mu \xi^\dag -\xi^\dag D_\mu
\xi)=-{D_\mu\pi\over F_\pi} +O(\pi^3) \\
V_\mu &=&{1\over 2}(\xi D_\mu \xi^\dag +\xi^\dag D_\mu \xi)\ \ \ .
\end{eqnarray}

For the $\Delta$, we
use the isospurion formalism\cite{hhk}, treating the $\Delta$ field
$T_\mu^i(x) $ as
a vector spinor in both
spin and  isospin space with the constraint $\tau^i T_\mu^i
(x)=0$. The components of
this field are
\begin{equation}
T^3_\mu =-\sqrt{{2\over 3}}\left( \begin{array}{l} \Delta^+\\ \Delta^0
 \end{array} \right)_\mu\ \ \ , \ T^+_\mu =\left( \begin{array}{l}
\Delta^{++}\\
\Delta^+/\sqrt{3}
 \end{array} \right)_\mu \ \ \ , \ T^-_\mu =-\left( \begin{array}{l}
\Delta^0/\sqrt{3}\\
\Delta^-
 \end{array} \right)_\mu\ \ \ .
\end{equation}
The field $T^i_\mu$ also satisfies the constraints for the
ordinary Schwinger-Rarita spin-${3\over 2}$ field,
\begin{equation}
\gamma^\mu T_\mu^i=0\ \ \  {\hbox{and}}\ \ \  p^\mu T_\mu^i=0\ \ \ .
\end{equation}
We eventually convert to the heavy baryon expansion, in which case the
latter constraint
becomes $v^\mu T_\mu^i=0$ with $v_\mu$ the heavy baryon velocity.

The relativistic parity-conserving (PC) Lagrangian for $\pi$, $N$, $\Delta$
interactions needed here is
\begin{eqnarray}
\label{eq:lpc}\nonumber
{\cal L}^\sst{PC}&=&{F_\pi^2\over 4} Tr D^\mu \Sigma D_\mu \Sigma^\dag +
\bar N (i {\cal D}_\mu \gamma^\mu -m_N) N + g_A \bar N A_\mu \gamma^\mu
\gamma_5 N\\ \nonumber
&&-T_i^\mu [(i{\cal D}^{ij}_\alpha\gamma^\alpha -m_\Delta \delta^{ij}
)g_{\mu\nu}-
{1\over 4} \gamma_\mu \gamma^\lambda (i{\cal D}^{ij}_\alpha\gamma^\alpha
-m_\Delta \delta^{ij} )
\gamma_\lambda \gamma^\nu \\ \nonumber
&&+{g_1\over 2} g_{\mu\nu} A_\alpha^{ij} \gamma^\alpha \gamma_5
+{g_2\over 2} (\gamma_\mu A_\nu^{ij} +A_\mu^{ij} \gamma_\nu )\gamma_5
+{g_3\over 2} \gamma_\mu A_\alpha^{ij} \gamma^\alpha\gamma_5 \gamma_\nu]
T_j^\nu \\
&&+g_{\pi N\Delta} [\bar T^\mu_i (g_{\mu\nu} +z_0 \gamma_\mu\gamma_\nu)
\omega_i^\nu N +h.c.]\ \ \ ,
\end{eqnarray}
where $\omega_\mu^i={\rm tr}[\tau^iA_\mu]/2$ while
$D_\mu$ and ${\cal D}_\mu$ are the gauge and chiral covariant
derivatives, respectively.
Explicit expressions for the fields and the
transformation properties can be found in \cite{zhu}.  Here,
$z_0$ is an off-shell parameter, which is not relevant in the
present work \cite{hhk}.

In order to obtain proper chiral counting for the nucleon, we
employ the conventional heavy baryon expansion of
${\cal L}^\sst{PC}$, and in order to cosistently include the $\Delta$ we
follow the small scale expansion proposed in \cite{hhk}. In this
approach energy-momenta and the delta and nucleon mass difference $\delta$ are
both treated as small expansion parameters in chiral power counting. The
leading
order
vertices in this framework can be obtained via $P_+ \Gamma P_+$ where
$\Gamma$ is
the original vertex in the relativistic Lagrangian and
\begin{equation}
P_{\pm}={1\pm \notv\over 2}\ \ \ .
\end{equation}
are projection operators for the large, small components of the Dirac
wavefunction respectively.
We collect some of the relevant terms below:
\begin{eqnarray}\nonumber
{\cal L}^\sst{PC}_v &=& \bar N [iv\cdot D +2g_A S\cdot A]N
-i {\bar T}^\mu_i [iv\cdot D^{ij} -\delta^{ij} \delta +g_1 S\cdot A^{ij}]
T_\mu^j \\
&&+g_{\pi N\Delta}[{\bar T}^\mu_i \omega_\mu^i N
+ \bar N \omega_\mu^{i\dag} T_i^\mu]
\end{eqnarray}
where $S_\mu$ is the Pauli-Lubanski spin operator and  $\delta \equiv m_\Delta
-m_N$.

The PV analog of Eq. (\ref{eq:lpc}) can be constructed using the chiral fields
$X^a_{L,R}$ defined as \cite{kaplan}:
\begin{equation}
X_L^a=\xi^\dag \tau^a \xi \  ,\ \    X_R^a=\xi \tau^a \xi^\dag \  ,\ \
X_{\pm}^a = X_L^a {\pm} X_R^a\ \ \ .
\end{equation}
We find it convenient to
follow the convention in Ref. \cite{kaplan} and separate the PV Lagrangian
into
its various isospin components.

The hadronic weak interaction has the form
\begin{equation}\label{38}
{\cal H}_\sst{W} = {G_\mu\over\sqrt{2}}J_\lambda J^{\lambda\ \dag}\ + \
{\hbox{h.c.}}\ \ \ ,
\end{equation}
where $J_\lambda$ denotes either a charged or neutral weak current built
out of quarks. In the Standard Model, the strangeness conserving
charged currents are pure isovector, whereas the
neutral currents contain both isovector and isoscalar components.
Consequently, ${\cal H}_\sst{W}$ contains $\Delta I=0, 1, 2$ pieces and
these channels
must all be accounted for in any realistic hadronic effective theory.

We quote here the relativistic Lagrangians, but employ the heavy baryon
projections,
as described above, in computing loops. It is straightforward to
obtain the corresponding heavy baryon Lagrangians from those listed below,
so we do not list the PV heavy baryon terms below. For the $\pi N$ sector
we have
\begin{eqnarray}\label{n1}
{\cal L}^{\pi N}_{\Delta I=0} &=&h^0_V \bar N A_\mu \gamma^\mu N \\
\nonumber && \\
\label{n2}
{\cal L}^{\pi N}_{\Delta I=1} &=&{h^1_V\over 2} \bar N  \gamma^\mu N  Tr
(A_\mu X_+^3)
-{h^1_A\over 2} \bar N  \gamma^\mu \gamma_5N  Tr (A_\mu X_-^3)\\
\nonumber
&& \ \ \ -{h_{\pi}\over 2\sqrt{2}}F_\pi \bar N X_-^3 N\\
\nonumber && \\
\label{n3}
{\cal L}^{\pi N}_{\Delta I=2} &=&h^2_V {\cal I}^{ab} \bar N
[X_R^a A_\mu X_R^b +X_L^a A_\mu X_L^b]\gamma^\mu N \\ \nonumber
&& \ \ \ -{h^2_A\over 2} {\cal I}^{ab} \bar N
[X_R^a A_\mu X_R^b -X_L^a A_\mu X_L^b]\gamma^\mu\gamma_5 N \; ,
\end{eqnarray}
where ${\cal I}^{ab}$ is a matrix coupling the $X^{a,b}$ to $I=2, I_3=0$.
The above Lagrangian was first given by Kaplan and Savage (KS)\cite{kaplan}.
However, the coefficients used in our work are
slightly different from those of Ref. \cite{kaplan} since our definition of
$A_\mu$ differs by an overall phase.

The term proportional to $h_\pi$ contains no derivatives. At
leading-order in $1/F_\pi$, it
yields the PV $NN\pi$ Yukawa coupling traditionally used in meson-exchange
models for the PV NN interaction \cite{ddh,haxton}.
Unlike the PV Yukawa interaction, the vector and axial vector terms in Eqs.
(\ref{n1}-\ref{n3}) contain derivative  interactions. The terms containing
$h_A^1$ and  $h_A^2$
start off with $NN\pi\pi$ interactions, while all the other terms start off
as $NN\pi$.
Such derivative interactions have not been included in conventional analyses
of nuclear and hadronic PV experiments. Consequently, the experimental
constraints on the low-energy constants $h_V^i$, $h_A^i$ are unknown.

It is useful to list the first few terms obtained by expanding the
Lagrangians in
Eqs. (\ref{n1}-\ref{n3}) in $1/F_\pi$. For the present purposes, the
following terms are needed:
\begin{eqnarray}\label{pi-n-n}
{\cal L}^{\pi NN}_{\hbox{Yukawa}} =-ih_{\pi} (\bar p n \pi^+ -\bar n p \pi^-)
[1-{1\over 3F_\pi^2} (\pi^+\pi^- +{1\over 2} \pi^0\pi^0)]&\\
{\cal L}^{\pi NN}_{V} =-{h_V^0+4/3 h_V^2\over \sqrt{2}F_\pi} [ \bar p\gamma^\mu
n D_\mu\pi^+ + \bar n \gamma^\mu p D_\mu \pi^-] &\\ \nonumber
{\cal L}^{\pi NN}_{A} = i{h_A^1+h_A^2\over F^2_\pi}\bar p\gamma^\mu \gamma_5 p
(\pi^+ D_\mu \pi^--\pi^- D_\mu \pi^+) &\\ \nonumber
+i{h_A^1-h_A^2\over F^2_\pi}\bar n\gamma^\mu \gamma_5 n
(\pi^+ D_\mu \pi^--\pi^- D_\mu \pi^+) &\\
+i{\sqrt{2}h_A^2\over F^2_\pi}\bar p\gamma^\mu \gamma_5 n \pi^+ D_\mu \pi^0
-i{\sqrt{2}h_A^2\over F^2_\pi}\bar p\gamma^\mu \gamma_5 n \pi^+ D_\mu \pi^0
&\; .
\end{eqnarray}
For the PV $\pi NN$ Yukawa coupling we have also kept terms with three pions.

The corresponding PV Lagrangians involving a $N\to\Delta$ transition are
somewhat more complicated. We relegate the complete expressions to Appendix
A, and
give here only the leading terms required for our calculation. As noted in
Ref.
\cite{zhu}, the one-pion $\pi N\Delta$ PV Lagrangian vanishes at leading order
in the heavy baryon expansion. The two-pion terms are
\begin{eqnarray}\label{pi-n-d}\nonumber
{\cal L}_A^{\pi N\Delta} =-{i h_A^{p \Delta^{++} \pi^- \pi^0}\over F_\pi^2}
\bar p \Delta_\mu^{++} D^\mu \pi^- \pi^0
-{i h_A^{p \Delta^{++} \pi^0 \pi^-}\over F_\pi^2}
\bar p \Delta_\mu^{++} D^\mu \pi^0 \pi^- &\\ \nonumber
-{i h_A^{p \Delta^{+} \pi^0 \pi^0}\over F_\pi^2}
\bar p \Delta_\mu^{+} D^\mu \pi^0 \pi^0
-{i h_A^{p \Delta^{+} \pi^+ \pi^-}\over F_\pi^2}
\bar p \Delta_\mu^{+} D^\mu \pi^+ \pi^-& \\ \nonumber
-{i h_A^{p \Delta^{+} \pi^- \pi^+}\over F_\pi^2}
\bar p \Delta_\mu^{+} D^\mu \pi^- \pi^+
-{i h_A^{p \Delta^{0} \pi^+ \pi^0}\over F_\pi^2}
\bar p \Delta_\mu^{0} D^\mu \pi^+ \pi^0 &\\ \nonumber
-{i h_A^{p \Delta^{0} \pi^0 \pi^+}\over F_\pi^2}
\bar p \Delta_\mu^{0} D^\mu \pi^0 \pi^+
-{i h_A^{p \Delta^{-} \pi^+ \pi^+}\over F_\pi^2}
\bar p \Delta_\mu^{-} D^\mu \pi^+ \pi^+ &\\ \nonumber
-{i h_A^{n \Delta^{++} \pi^- \pi^-}\over F_\pi^2}
\bar n \Delta_\mu^{++} D^\mu \pi^- \pi^-
-{i h_A^{n \Delta^{+} \pi^- \pi^0}\over F_\pi^2}
\bar n \Delta_\mu^{+} D^\mu \pi^- \pi^0 & \\ \nonumber
-{i h_A^{n \Delta^{+} \pi^0 \pi^-}\over F_\pi^2}
\bar n \Delta_\mu^{+} D^\mu \pi^0 \pi^-
-{i h_A^{n \Delta^{0} \pi^0 \pi^0}\over F_\pi^2}
\bar n \Delta_\mu^{0} D^\mu \pi^0 \pi^0 &\\ \nonumber
-{i h_A^{n \Delta^{0} \pi^+ \pi^-}\over F_\pi^2}
\bar n \Delta_\mu^{0} D^\mu \pi^+ \pi^-
-{i h_A^{n \Delta^{0} \pi^- \pi^+}\over F_\pi^2}
\bar n \Delta_\mu^{0} D^\mu \pi^- \pi^+ & \\
-{i h_A^{n \Delta^{-} \pi^+ \pi^0}\over F_\pi^2}
\bar n \Delta_\mu^{-} D^\mu \pi^+ \pi^0
-{i h_A^{n \Delta^{-} \pi^0 \pi^+}\over F_\pi^2}
\bar n \Delta_\mu^{-} D^\mu \pi^0 \pi^+\ \ + {\hbox{h.c.}} &\; ,
\end{eqnarray}
where the couplings $h_A^{p \Delta^{++} \pi^- \pi^0}$ are defined in terms
of the
various SU(2) PV low-energy constants in Appendix A.

The PV $\pi\Delta\Delta$ Lagrangians, also listed in Appendix A, contain
terms analogous
to the Yukawa, $V$, and $A$ terms in Eqs. (\ref{n1}-\ref{n3}). Since we
compute corrections
up to one-loop order only, and since the initial and final states are
nucleons, the PV $\pi\pi
\Delta\Delta$ terms ($A$-type) are not relevant here. The leading,
single-$\pi$ Yukawa and $V$-type
interactions are
\begin{eqnarray}\nonumber
{\cal L}^{\pi \Delta \Delta}_{\hbox{Yukawa}}=-i{h_\Delta\over \sqrt{3}}
(\bar \Delta^{++}\Delta^+\pi^+ -\bar \Delta^{+}\Delta^{++}\pi^- ) & \\
\nonumber
-i{h_\Delta\over \sqrt{3}}
(\bar \Delta^{0}\Delta^-\pi^+ -\bar \Delta^{-}\Delta^{0}\pi^- ) & \\
-i{2h_\Delta\over 3}
(\bar \Delta^{+}\Delta^0\pi^+ -\bar \Delta^{0}\Delta^{+}\pi^- ) & \\ \nonumber
{\cal L}_V^{\pi \Delta\Delta}=-{h_V^{\Delta^{++}\Delta^+}\over F_\pi}
( \bar \Delta^{++} \gamma_\mu \Delta^+ D^\mu \pi^+ +
\bar \Delta^{+} \gamma_\mu \Delta^{++} D^\mu \pi^- ) &\\ \nonumber
-{h_V^{\Delta^{+}\Delta^0}\over F_\pi}
( \bar \Delta^{+} \gamma_\mu \Delta^0 D^\mu \pi^+ +
\bar \Delta^{0} \gamma_\mu \Delta^{+} D^\mu \pi^- ) &\\
-{h_V^{\Delta^{0}\Delta^-}\over F_\pi}
( \bar \Delta^{0} \gamma_\mu \Delta^- D^\mu \pi^+ +
\bar \Delta^{-} \gamma_\mu \Delta^{0} D^\mu \pi^- ) &
\end{eqnarray}
where the coefficients are given in Appendix A.

One may ask whether there exist additional PV effective interactions that could
contribute at the order to which we work. In the pionic sector there exists
one CP-conserving, PV Lagrangian:
\begin{equation}
\label{eq:pionicpv}
{\cal L}^\sst{PV}_\pi = \epsilon_{ijk}  \omega_\mu^i \omega_\nu^j
(D^\mu \omega^\nu_k -D^\nu \omega_k^\mu )\ \ \ .
\end{equation}
At leading order in $1/F_\pi$, ${\cal L}_\pi$ contains five pions. Its
lowest order
contribution appears at two-loop order at best, so we do not consider it here.

Similarly, one may consider possible contributions from two-derivative
operators. There
exists one CP-conserving, PV operator:
\begin{equation}
\label{eq:twoderivpv}
{1\over \Lambda_\chi}\bar N \sigma^{\mu\nu} [D_\mu A_\nu -D_\nu A_\mu] N\ \ \ .
\end{equation}
There exist three independent PC, two-derivative operators \cite{Fet98}. For
example, one may choose the following three:
\begin{equation}
{1\over \Lambda_\chi}\bar N i\gamma_5 D_\mu A^\mu N\; ,
\end{equation}
\begin{equation}
{1\over \Lambda_\chi}\bar N A^\mu A_\mu N\; ,
\end{equation}
\begin{equation}\label{eq:twoderivpc}
{1\over \Lambda_\chi}\bar N \sigma^{\mu\nu} [A_\mu, A_\nu] N\; .
\end{equation}
As we discuss in Appendix B, none of the two-derivative operators in Eqs.
(\ref{eq:twoderivpv}) -
(\ref{eq:twoderivpc}) contribute to the renormalization of $h_\pi$ at the
order to which we work in the present analysis.

\section{The loop corrections}
\label{sec3}

The leading order loop corrections to the Yukawa interaction of Eq.
(\ref{pi-n-n}) are
generated by the diagrams of Figs. 1-2. As we discuss in Appendix B, the
contributions from
many of the diagrams which nominally renormalize $\hpi$ vanish at the order
at which
we truncate. In particular, none of the vector ($V$-type) $\pi NN$ and
$\pi\Delta\Delta$ terms contribute
to this order. In what follows, we discuss only the non-vanishing Yukawa
and $A$-type contributions.
Details regarding the vanishing of the other contributions appear in
Appendix B. Following
the conventional practice, we regulate the loop integrals using dimensional
regularization. The
pole terms proportional to $1/D-4$ are cancelled by appropriate
counterterms. We identify only
the terms nonanalytic in quark masses with the loops. All other analytic
terms are indistinguishable
from finite parts of the corresponding counterterms.

The nonvanishing contribution from Fig. 1(a) arises from the insertion of
the $3\pi$ part of
the Yukawa interaction of Eq. (\ref{n2}). The nonanalytic term is
\begin{equation}\label{m_a}
iM_{(a)}={5\over 6}{m_\pi^2\over \Lambda_\chi^2}\ln ({\mu\over m_\pi})^2
h_\pi \tau^+ \; ,
\end{equation}
where $\Lambda_\chi =4\pi F_\pi$ and $\mu$ is the subtraction scale
introduced in dimensional
regularization. For simplicity, we show here only the contributions for
$n\to p\pi^-$. The
terms for $p\to n\pi^+$ are equal in magnitude and opposite in sign
since it is the hermitian conjugate of the $n\to p\pi^-$ piece. This property
holds to all orders of chiral expansion.

The nonvanishing contribution from Fig. 1 (b) arises from strong vertex
correction
to the leading order $\pi NN$ Yukawa interaction:
\begin{equation}\label{m_b}
iM_{(b)}={3\over 4}g_A^2{m_\pi^2\over \Lambda_\chi^2}\ln ({\mu\over
m_\pi})^2 h_\pi \tau^+ \; .
\end{equation}
The terms in Figs. 1(c1)-(c2) are generated by the PV axial $\pi\pi NN$
couplings
proportional to the $h_A^i$. We have
\begin{equation}\label{m_c}
iM_{(c1)+(c2)}=2\sqrt{2}\pi g_A{m_\pi^3\over F_\pi \Lambda_\chi^2} h_A^1 \tau^+
\; .
\end{equation}
The contribution from $h_A^2$ to these two diagrams cancels out, leaving
only the dependence on $h_A^1$.
We note that although this term is propotional to $m_q^{3/2}$ and, thus,
nominally suppressed, the
coefficient of $h_A^1$ is fortuitously large ($\sim 1/4$).
The two pion vertex in FIG. 1 (d1)-(d2) comes from the chiral connection
$V_\mu$:
\begin{equation}\label{m_d}
iM_{(d1)+(d2)}=-{1\over 2}{m_\pi^2\over \Lambda_\chi^2}\ln ({\mu\over
m_\pi})^2 h_\pi \tau^+ \; .
\end{equation}

The leading contribution involving $\Delta$ intermediate states arises from
Fig. 2(a).
The corresponding amplitude receives contributions from
three different isospin combinations for the $\Delta$
intermediate states. Their sum reads
\begin{equation}\label{m_2a}
iM_{2(a)}=-{20\over 9}{g_{\pi N\Delta}^2 h_\Delta\over \Lambda_\chi^2}
[(2\delta^2-m_\pi^2) \ln ({\mu\over m_\pi})^2 -4\delta \sqrt{\delta^2-m_\pi^2}
\ln {\delta +\sqrt{\delta^2-m_\pi^2}\over m_\pi} ] \tau^+ \; .
\end{equation}
The corrections generated by the  PV $\pi\pi N\Delta$ vertices are
\begin{equation}\label{m_2b}
iM_{2(b1)+2(b2)}={2\over 3}{g_{\pi N\Delta} \over F_\pi \Lambda_\chi^2}
[(\delta^2-{3\over 2}m_\pi^2)\delta \ln ({\mu\over m_\pi})^2 -2
(\delta^2-m_\pi^2)^{3/ 2}
\ln {\delta +\sqrt{\delta^2-m_\pi^2}\over m_\pi} ] h_A^\Delta \tau^+  \; ,
\end{equation}
where $h_A^\Delta$ is defined as
\begin{equation}
\label{eq:hadelta}
h_A^\Delta ={1\over
\sqrt{3}}(h_A^{n\Delta^0\pi^+\pi^-}+h_A^{p\Delta^+\pi^-\pi^+})
+\sqrt{2\over 3} (h_A^{n\Delta^+\pi^0\pi^-}-h_A^{p\Delta^0\pi^0\pi^+})
-h_A^{n\Delta^{++}\pi^-\pi^-}-h_A^{p\Delta^-\pi^+\pi^+}\; .
\end{equation}

Summing all the non-vanishing loop contributions yields the following
expression for $\Delta\hpi$:
\begin{eqnarray}
\label{eq:deltahpi}\nonumber
\Delta\hpi& = &{1\over 3}{m_\pi^2\over \Lambda_\chi^2}\ln ({\mu\over
m_\pi})^2 h_\pi
+{3\over 4}g_A^2{m_\pi^2\over \Lambda_\chi^2}\ln ({\mu\over
m_\pi})^2 h_\pi+2\sqrt{2}\pi g_A{m_\pi^3\over F_\pi \Lambda_\chi^2} h_A^1
\\  \nonumber
&&-{20\over 9}{g_{\pi N\Delta}^2 h_\Delta\over \Lambda_\chi^2}
[(2\delta^2-m_\pi^2) \ln ({\mu\over m_\pi})^2 -4\delta \sqrt{\delta^2-m_\pi^2}
\ln {\delta +\sqrt{\delta^2-m_\pi^2}\over m_\pi} ]\\ &&
+{2\over 3}{g_{\pi N\Delta} \over F_\pi \Lambda_\chi^2}
[(\delta^2-{3\over 2}m_\pi^2)\delta \ln ({\mu\over m_\pi})^2 -2
(\delta^2-m_\pi^2)^{3/ 2}
\ln {\delta +\sqrt{\delta^2-m_\pi^2}\over m_\pi} ] h_A^\Delta
\end{eqnarray}

The final nonvanishing corrections arise from $N$ and $\pi$ wavefunction
renormalization.
These corrections, which have been computed previously \cite{npa},
generate deviations from
unity of $Z_N$ and $\sqrt{Z_\pi}$ appearing in the expression for $\hpieff$
in Eq. (\ref{eq:hpieff}).
In the case of $Z_N$, the nonvanishing contributions arise from Figs.
1(e1)-(e2) and
2(c1)-(c2):
\begin{eqnarray}\label{z-n}\nonumber
Z_N-1=
 {9\over 4} g_A^2 {m_\pi^2\over \Lambda_\chi^2} \ln ({\mu\over m_\pi})^2
-4 g_{\pi N\Delta}^2 [ {2\delta^2- m_\pi^2\over \Lambda_\chi^2}\ln
({\mu\over m_\pi})^2 &\\
\label{eq:nwaveren}
-4{\delta\sqrt{ \delta^2-m_\pi^2} \over \Lambda_\chi^2} \ln {\delta
+\sqrt{\delta^2-m_\pi^2}\over
m_\pi} ]
\ \ \ .
\end{eqnarray}
The pion's wavefunction renormalization arises from Fig. 2 (k) \cite{book}:
\begin{equation}
\label{eq:piwaveren}
\sqrt{Z_\pi}-1 = -{1\over 3}\left({m_\pi\over\Lambda_\chi}\right)^2
\ln\left({\mu\over m_\pi}\right)^2\
\ \ .
\end{equation}
Numercially, the loop contributions to $\sqrt{Z_\pi}$ are small compared to
those entering $Z_N$.

Note the one loop renormalization of $h_\pi$ from the PV Yukawa $\pi NN$ and
$\pi\Delta\Delta$ vertices is already at the order $1/\Lambda_\chi^2$. An
additional loop
will introduce a factor of $1/\Lambda_\chi^2$. Loops containing the axial
vector
$NN\pi\pi$ and $N\Delta\pi\pi$ vertices and one strong $NN\pi$ or
$N\Delta\pi$ vertex
are of ${\cal O}(1/\lamchis F_\pi)$. To obtain contributions of ${\cal
O}(1/\lamchic)$,
one would require the insertion of operators carrying explicit factors of
$1/\lamchi$
into one loop graphs. We find no such contributions.

\section{The scale of loop corrections}
\label{sec5}
We may estimate the numerical importance of the loop corrections to
$\hpibare$ by
taking $\delta =0.3$ GeV, $g_A=1.267$ \cite{pdg} and $g_{\pi N\Delta}=1.05$
\cite{hhk} and by choosing $\mu=\Lambda_\chi =1.16$ GeV\footnote{Since the
dependence on $\mu$ is logarithmic, one may choose other values, such as
$\mu=m_\rho$, without affecting the numerical results significantly}. With
these
inputs, the value of $Z_N\sqrt{Z_\pi}$ is completely determined. The vertex
corrections,
which appear as $\Delta\hpi$ in Eq. (\ref{eq:hpieff}), depend on the PV
couplings
$\hpi$, $h_A^1$, $h_\Delta$, and $h_A^\Delta$. We obtain
\begin{equation}\label{eff}
\hpieff=0.5 h_\pi  +0.25 h_A^1 -0.24 h_\Delta +0.079 h_A^\Delta \; .
\end{equation}
Note that the effect of the wavefunction renormalization corrections is to
reduce
the dependence on $\hpibare$ by roughly 50\%. In addition, the dependence of
$\hpieff$ on $h_A^1$ and $h_\Delta$ is non-negligible. Their coefficients are
only a factor of two smaller than that of $\hpibare$. Although these
contributions
arise at ${\cal O}(p^2,\ p^3)$, they are fortuitously enhanced numerically.
Thus,
in a complete
anaysis of the OPE PV interaction one should not ignore these constants.

At present, one has no direct experimental constraints on the parameters
$h_A^1$, $h_\Delta$,
and $h_A^\Delta$, as a comprehensive analysis of hadronic PV data including
the full chiral
structure of the PV hadronic interaction has yet to be performed.
Consequently, one must rely
on theoretical input for guidance regarding the scale of the unknown
constants.
Estimates of $h_A^1$ are given by the authors of Ref. \cite{kaplan}. These
authors observe
that the usual pole dominance approximation for P-wave non-leptonic hyperon
decays typically
underpredicts the experimental amplitudes by a factor of two. The
difference may be resolved
by the inclusion of local, parity-conserving operators having structures
analogous to the
$A$-type terms in Eq. (\ref{n2}). The requisite size of the $\Delta S=1$
contact terms may
imply a scale for the analogous $\Delta I=1$ PV terms. If so, one might
conclude that
$h_A^1$ should be on the order of $10 g_\pi$. On the other hand, a simple
factorization estimate
leads to $h_A^1\sim 0.2 g_\pi$. While the sign of $h_A^1$ is fixed in the
factorization approximation,
the sign of the larger value is undetermined. Thus,
it is reasonable to conclude that $h_A^1$ may
be large enough to significantly impact $\hpieff$, though considerably more
analysis is needed to
yield a firm conclusion.

The $\pi\Delta\Delta$ Yukawa coupling $h_\Delta$ has been estimated in Ref.
\cite{fcdh} using methods
similar to those of Ref. \cite{ddh}. The authors quote a \lq\lq best value"
of $h_\Delta= -20 g_\pi$,
with a \lq\lq reasonable range" of $(-51\to 0)\times g_\pi$.\footnote{This
coupling is denoted
$f_{\Delta\Delta\pi}$ in Ref. \cite{fcdh}.} Na\"\i vely, subsitution of the
best value into Eq.
(\ref{eff}) would increase the value of $\hpieff$, whereas the $^{18}$F
result would seem to
require a reduction. As we argue below, however, the relationship between
the couplings computed
in Refs. \cite{ddh,fcdh} and the parameters appearing in Eq. (\ref{eff}) is
somewhat ambiguous.
Direct substitution of the theoretical value into $\hpieff$ may not be
entirely appropriate.

To date, no theoretical estimate of the $A$-type $\pi\pi N\Delta$ coupling
has been performed. A simple estimate of the scale is readily obtained
using the
factorization approximation. To that end, we work with tree-level form
of $\hweak$. Neglecting short-distance QCD corrections and terms containing
strange quarks, one has
\begin{eqnarray}
\label{eq:hwtree}
\hweak(\Delta S=0) & = & {G_F\over\sqrt{2}} \big\{ \cos^2\theta_c {\bar
u}\gamma_\lambda(1-\gamma_5) d {\bar d}\gamma^\lambda(1-\gamma_5)u \\
&& -2(1-2\sstw)V_\lambda^{(3)} A^{(3)\ \lambda}+\frac{4}{3}\sstw
  V_\lambda^{(0)} A^{(3)\ \lambda}\big\} \ \ \ ,
\end{eqnarray}
where $V_\lambda^{(3)}$ and $A_\lambda^{(3)}$ denote the third components
of the octet of vector and axial vector currents, respectively, and
\begin{equation}
V_\lambda^{(0)}=\frac{1}{2} ({\bar u}\gamma_\lambda u+ {\bar
d}\gamma_\lambda d)
\ \ \ .
\end{equation}
Consider now the first term in the expression for $h_A^\Delta$ given in
Eq. (\ref{eq:hadelta}). In the
factorization approximation,
$\hweak$ contributes only to the antisymmetric combination
\begin{equation}
\frac{1}{2}(h_A^{n\Delta^0\pi^+\pi^-} - h_A^{n\Delta^0\pi^-\pi^+})\ \ \ .
\end{equation}
The neutral current contribution to this combination, which arises only from
the term containing $V_\lambda^{(3)}$, is
\begin{equation}
\sqrt{2} G_F F_\pi^2 (1-2\sstw) C_5^A(n\Delta^0)\approx 2 g_\pi
C_5^A(n\Delta^0)
\ \ \ ,
\end{equation}
where $C_5^A(n\Delta^0)\sim{\cal O}(1)$ is the axial vector $n\to\Delta^0$
form factor at the photon point. After Fierz re-ordering, the charged current
component of $\hweak$ contributes roughly
\begin{equation}
-(4 g_\pi/3) C_5^A(n\Delta^0),
\end{equation}
yielding a total factorzation contribution of about $ (2 g_\pi/3)
C_5^A(n\Delta^0)$. Thus, one would expect the scale of the axial vector
$\pi\pi N\Delta$ couplings to be on the order of a few $\times\ g_\pi$.

In the
particular case of the combination appearing in $h_A^\Delta$, however, the
sum of factorization contributions cancels identically. As one sees from the
expressions for the $h_A^{N\Delta\pi\pi}$ given in Appendix A, isospin
requires
\begin{equation}
h_A^{n\Delta^0\pi^+\pi^-}+h_A^{p\Delta^+\pi^-\pi^+} = 0\ \ \ .
\end{equation}
The factorization contributions independently satisfy this sum rule. The
second combination of constants appearing in Eq. (\ref{eq:hadelta}),
\begin{equation}
h_A^{n\Delta^+\pi^0\pi^-}-h_A^{p\Delta^0\pi^0\pi^+}\ \ \ ,
\end{equation}
also vanishes in the factorization approximation, even though the individual
couplings do not. The third pair of couplings received no factorization
contributions. Thus, one has $h_A^\Delta=0$ in this approximation. In
principle,
non-factorization contributions yield a non-zero value for $h_A^\Delta$.
Although
we have not evaluated these contributions, we do not expect the scale to be
significantly larger than the factorization value for the individual
$h_A^{N\Delta\pi\pi}$ couplings. Consequently, we estimate a reasonable range
for $h_A^\Delta$ of $( 0 \to {\hbox{few}})\times g_\pi$.

These theoretical estimates suggest considerable ambiguity in the
prediction for $\hpieff$. In
principle, some of this ambiguity might be removed by performing the
comprehensive analysis of
hadronic PV suggested above, in which the various constants would be
determined entirely by experiment.
The viability of such a program remains to be seen.

\section{Comparing with Microscopic Calculations}
\label{sec6}

The results in Eqs. (\ref{eq:deltahpi}-\ref{eq:piwaveren}) embody the full
SU(2)
chiral structure at ${\cal O}(p^3)$ of $\bra{N\pi}\hweak\ket{N}$ at leading
order in the pion momentum. Any microscopic calculation of this matrix
element which respects the
symmetries of QCD should display the dependence on light quark masses
appearing in $\hpieff$. In
principle, an unquenched lattice QCD calculation with light quarks would
manifest this chiral
structure. In practice, however, unquenched calculations remain difficult,
and even quenched
calculations require the use of heavy quarks. For a lattice determination
of $\bra{N\pi}\hweak\ket{N}$,
the expressions in Eqs. (\ref{eq:deltahpi}-\ref{eq:piwaveren}) could be
used to extrapolate to the
light quark limit, much as the chiral structure of baryon mass and magnetic
moment
can be used for similar extrapolations \cite{thomas}.

In the absence of a first principles QCD calculation, one must rely on
symmetries and/or models
to obtain the PV $NN\pi$ coupling. A variety of such approaches have been
undertaken, including
the SU(6)$_w$/quark model calculation of Refs. \cite{ddh,fcdh}, the Skyrme
model \cite{t1},
and QCD sum rules \cite{t2}. To date, the DDH/FCDH analysis remains the most
comprehensive and has become the benchmark for comparison
between experiment and theory. Consequently, we focus on this work as a
\lq\lq case study" in the
problem of matching microscopic calculations onto hadronic effective theory.

The DDH/FCDH approach relies heavily on symmetry methods in order
to relate the PV $\Delta S=0$ matrix
elements to experimental $\Delta S=1$ nonleptonic hyperon decay
amplitudes.  All the
charged current (CC) contributions to the $\Delta S=0,1$ $B\to B' M$
amplitudes, where $M$ is a pseudoscalar meson, can be
related using SU(3) arguments.  Likewise, the neutral current (NC)
component of the effective weak Hamiltonian belonging
to the same multiplets as the CC components ({\it i.e.} those arising
from a product of purely left-handed currents) can also be related via
SU(3). The remaining NC
contributions to the $\Delta S=0$ PV amplitudes are computed using
factorization and the MIT bag model. The DDH approach also employs
SU(6)$_w$ symmetry
arguments in order to calculate parity-violating vector meson couplings.
Although
one requires only SU(3) to determine the pseudoscalar couplings,
we refer below to the general SU(6)$_w$ formalism used in Refs.
\cite{ddh,fcdh}.

The general SU(6)$_w$ analysis employed by DDH/FCDH introduces five reduced
matrix elements: $a_{t,v}$, $b_{t,v}$, and
$c_v$. These constants correspond to SU(6)$_w$ components of
the weak Hamiltonian:
\begin{equation}
[({\bar B}B)_{35}\otimes M_{35}]_{35}  \sim  c_v
\end{equation}
\begin{equation}
\label{eq:su6reduced}
[({\bar B}B)_{405}\otimes M_{35}]_{280,\overline{280}}  \sim  b_t, b_v
\end{equation}
\begin{equation}
[({\bar B}B)_{405}\otimes M_{35}]_{280,\overline{280}}  \sim  a_t, a_v
\end{equation}
One may represent these different components of $\hweak$ diagramatically as
in Fig. 3. The
components shown in Fig. 3a,b correspond to $b_{t,v}$ and $c_v$,
respectively. In practice,
these contributions are determined entirely from empirical
hyperon decay data. The term in Fig. 3a corresponds
to $a_{t,v}$ and is computed in Refs. \cite{ddh,fcdh} using factorization.

The PV $NN\pi$ Yukawa coupling can be expressed in terms of these SU(6)$_w$
reduced matrix elements plus an additional factorization/quark model term.
Temporarily neglecting short-distance QCD corrections to
$\hweak$, one has
\begin{eqnarray}
\label{eq:su6bare}
\bra{p\pi^-}\hweak\ket{n} & = & {1\over 3\sqrt{2}}\tan\theta_c c_v \\
  & - & {2\over 9\sqrt{2}} \csc 2\theta_c \sstw (2c_v-b_t) +{1\over 3}\sstw
y\ \ \ , \nonumber
\end{eqnarray}
where $\theta_c$ and $\theta_\sst{W}$ are the Cabibbo and Weinberg angles,
respectively, and $y$ denotes
a Fierz/factorization contribution. The first term on the RHS of Eq.
(\ref{eq:su6bare}) gives the
CC contribution, while the remaining terms arise from weak NC's. Including
short-distance QCD
renormalization of $\hweak$ leads to a modification of Eq. (\ref{eq:su6bare}):
\begin{eqnarray}
\bra{p\pi^-}\hweak\ket{n}& = & \left\{ [1-2\sstw]\gamma(K)
+\sin^2\theta_c\right\}{\rho\over
   \sin^2\theta_c} g_\pi \nonumber \\
\label{eq:su6qcd}
&& + \sin^2\theta_c (B_1+B_2)\ \ \ ,
\end{eqnarray}
where
\begin{eqnarray}
g_\pi& = & {1\over 3\sqrt{2}}\tan\theta_c c_v \\
B_1 & = & {4\over9\sqrt{2}}\eta E(K)
\left({1\over\sin\theta_c\cos\theta_c}\right)
   (b_v/6 -b_t/12-c_v/2) \\
B_2 & = & {1\over 3}F(K) y \ \ \ ,
\end{eqnarray}
and $\gamma(K)$, $E(K)$ and $F(K)$ are summed leading log
(renormalization group) factors dependent on
\begin{equation}
K = 1 - {\alpha_s(\mu)\over\pi} [11-{2\over 3} N_f] \ln{\mws\over\mu^2} \ \ \ .
\end{equation}
The overall scale factor $\rho$ appearing in Eq. (\ref{eq:su6qcd})
was introduced in Ref.\cite{ddh} in order
to account for various theoretical uncertainties entering the analysis.

The appearance of $c_v$, $b_t$, and $b_v$ in $g_\pi$ and $B_1$ relies on
{\em tree-level} SU(6)$_w$ symmetry---long-distance chiral corrections of
the types shown in Fig. 4 have not been explicitly included.
Inclusion of such corrections would necessitate a reanalysis
of the $\Delta S=1$ amplitudes in much the same way that one treats the
octet of baryon axial vector
currents \cite{j1} or magnetic moments \cite{mm}. For example, letting
$A(\Lambda^0_{-})$ denote
the amplitude for $\Lambda\to p\pi^-$ one has at tree-level
\begin{equation}
A(\Lambda^0_-) = {1\over\sqrt{3}}(b_v/6-b_t/12-c_v/2)\ \ \ .
\end{equation}
Including the leading chiral corrections would yield the modification
\begin{equation}
A(\Lambda^0_-) = {1\over\sqrt{3}}\sqrt{Z_\Lambda Z_p Z_\pi}
(b_v/6-b_t/12-c_v/2)
+\Delta A(\Lambda^0_-)\ \ \ ,
\end{equation}
where $\Delta A(\Lambda^0_-)$ denotes vertex corrections and possible
contributions from higher-dimension
operators. Similar corrections would appear in the SU(6)$_w$ symmetry terms
in Eqs. (\ref{eq:su6bare},
\ref{eq:su6qcd}). Given the absence of these corrections from the DDH/FCDH
analysis, the symmetry
components $\bra{p\pi^-}\hweak\ket{n}$ do not formally embody the
subleading chiral structure of
$\hpieff$.  The {\em numerical} impact of applying chiral corrections to
the DDH/FCDH tree-level
SU(6)$_w$ analysis is much less clear, since some of the chiral
modifications can be absorbed into renormalized values of
the chiral couplings, which are determined empirically.  Nevertheless,
the potentially sizeable effect of the SU(2) chiral
corrections on $\hpieff$ should give one pause.

A related issue is the
degree to which ambiguities introduced by
kaon and $\eta$ loops in SU(3) HBChPT could plague an analysis of
the $\Delta S=1$ amplitudes.   Here recent work by Donoghue and Holstein
argues that finite nucleon size call for long-distance regularization
of such heavy meson loops, which substantially reduces their effects\cite{dh}.
Results are then similar to what arises from use of a cloudy bag approach
to such matrix elements\cite{cb}.  A comprehensive study of
such issues -- and their impact on the DDH/FCDH calculation of $\hpi$ -- goes
beyond the scope of the present work. Nevertheless, the sizeable impact of
the chiral
corrections in $\hpieff$ and the use of tree-level symmetry arguments in
Refs. \cite{ddh,fcdh} points to a possibly significant mismatch
between $\hpieff$ and $h_\pi^\sst{DDH}$.

The remaining terms in the DDH/FCDH analysis -- involving the parameters
$\eta$ and $y$ -- are
determined by explicit MIT bag model calculations. One may ask whether the
latter effectively
includes any part of the subleading chiral structure of $\hpieff$. In order to
address this question, we make three observations:

\medskip

\noindent {\bf 1. Sea quarks and gluons generate $c_v$}. The parameter
$c_v$ vanishes
identically in any quark model in which baryons consist solely of three
constituent quarks.
The $\Delta S=1$ hyperon decay data, however,
clearly implies that $c_v\not= 0$.  In order to obtain a nonzero value in a
quark model, one
requires the presence of sea quarks and gluons. It is shown in \cite{gluon} for
example, that $c_v\not= 0$ when gluons are added to the MIT bag model.
Similarly, one would expect contributions from the $q\bar q$ pairs in the sea.
Since relativistic quark models already contain $q\bar q$ pairs in
the form of \lq\lq Z-graphs" \cite{Gei97}, it is likely that disconnected
$q\bar q$
insertions (see Fig. 5b)  give the dominant sea quark contribution to $c_v$.
In a chirally corrected analysis of nonleptonic
decays, the long-distance part of the disconnected $q\bar q$ insertions
appear explicitly in
the guise of pseudoscalar loops, while the short-distance contributions are
subsumed into the
value of $c_v$ and possible higher dimension operators. \lq\lq Quenched"
quark models without
explicit pionic degress of freedom generally do not include the long-distance
physics of disconnected
insertions.

\medskip


\medskip

\noindent{\bf 2. The $m_q$-dependence is different}. In conventional HBChPT
analyses of hadronic
observables, one only retains the loop contributions non-analytic in the
light quark mass. The
constituent quark model (without explicit pions) has a difficult time
producing these non-analytic
contributions. The simplest, illustrative example is the nucleon isovector
charge radius, $\langle
r^2\rangle_{T=1}$, which is singular in the chiral limit \cite{radius}. This
chiral
singularity, of the form $\ln m_\pi^2 \sim \ln m_q$, is produced by $\pi$
loops. Relativistic quark
models, such as the MIT bag model, yield a finite value for  $\langle
r^2\rangle_{T=1}$ as $m_q\to 0$.
One cannot produce the chiral singularity in a quark model without
including disconnected $q\bar q$
insertions dressed as mesons.

The corresponding argument in the case of $\hpieff$ is less direct, but
still straightforward.
In the limit of a degenerate $N$ and $\Delta$, the non-analytic terms in
$\hpieff$ have quark mass-dependences of the
form $m_q\ln m_q$ or $m_q^{3/2}$. As we show in Appendix C,
bag model matrix
elements of the four quark operators appearing in $\hweak$ have a Taylor
series expansion about
$m_q=0$. Thus, the parameters $\eta$ and $y$ cannot contain the
non-analytic structures generated
by the diagrams in Figs. 1-2.

\medskip

\noindent{\bf 3. Graphs are missing}. This observation is simply a
diagrammatic summary of the
previous two observations. For simplicity, consider a subset of the
quark-level diagrams
associated with the appearance of $h_A^i$ in $\hpieff$. Typical
contributions to the axial
$NN\pi\pi$ PV vertex are shown in Fig. 5a. The corresponding loop
contributions to $\hpieff$ appear
in Fig. 5b,c. Those in Fig. 5b involve disconnected $q\bar q$ insertions,
which do not occur in the
constituent quark model. The contribution of Fig. 5c involves Z-graphs,
which are produced in a relativistic quark model\footnote{e.g., as a
correction to
the $b_{t,v}$ terms of Fig. 3b}. In principle, the $3q+q\bar q$
intermediate state could
contain an $N\pi$ pair. As argued previously, however, the Z-graphs
implicit in the MIT bag model
calculation of $\hpi$ do not produce the nonanalytic structure of the
corresponding $\pi$ loop.
Apparently, only an unquenched quark model, which generates the
disconnected insertions of Fig. 5b,
could produce the requisite nonanalytic terms.

From this \lq\lq case study" of the DDH/FCDH calculation of $\hpi$, we
conclude that
the SU(6)$_w$/quark model approach used in Refs. \cite{ddh,fcdh} does not
incorporate the chiral
structure of $\hpieff$. Were the numerical impact of the chiral corrections
negligible, this
observation would not be bothersome. The actual impact of the chiral
corrections, however, is
significant. Thus, it is perhaps not surprising that the $^{18}$F result
and the DDH/FCDH \lq\lq best
value" do not agree.

\section{Conclusions}
\label{sec7}

With the confirmation of the electroweak sector of the  Standard Model at
the 1\% level or
better in a variety of leptonic and semi-leptonic processes, one has little
reason to doubt its
validity in the purely hadronic domain. Similarly, the predictions of QCD
in the perturbative regime
have been confirmed with a high degree of confidence. Thus, one may
justifiably consider $\hweak$,
the effective Hamiltonian including its perturbative strong interaction
correction, to be
well understood. Moreover, the precision available with present and future
hadronic PV experiments is
unlikely to match the levels achieved in leptonic and semileptonic
processes. Consequently, one has
little hope of detecting small deviations in $\hweak$ from its SM structure
due to \lq\lq new
physics".  On the other hand, much about QCD in the non-perturbative regime
remains mysterious:
the mechanism of confinement, the dynamics of chiral symmetry breaking, the
role or sea quarks in
the low-energy structure of the nucleon, and so forth. Each of these issues
bears on one's understanding
of {\em matrix elements} of $\hweak$. In this sense, the low-energy, PV
hadronic weak interaction
constitutes a probe of the dynamics of low-energy QCD, in a manner
analogous to the probe provided
by the electromagnetic interaction.

From a phenomenological standpoint, the matrix element one may hope to
extract from hadronic
PV observables with the least ambiguity is $\bra{N\pi}\hweak\ket{N}$. In
this study, we have
argued that any theoretical interpretation of this matrix element must take
into account the
consequences of chiral symmetry. Indeed the
chiral corrections to the tree-level, PV $\pi NN$ Yukawa
coupling are not small. At ${\cal O}(1/\lamchic)$, the effective coupling
measured in experiments,
$\hpieff$, depends not only on the bare coupling, $\hpibare$, but also on
new (and experimentally undetermined) PV low-energy constants,
$h_A^1$, $h_A^\Delta$, and $h_\Delta$, as well. Furthermore, the coefficients
of $\hpibare$, $h_A^1$,
and $h_\Delta$ are comparable in magnitude. At present, one has only simple
theoretical estimates
of the magnitudes of the $h_A^1$ and $h_A^\Delta$ in addition to the FCDH
calculation of
$h_\Delta$. These estimates suggest that the new PV couplings appearing in
$\hpieff$ could be
as large as $\hpibare$. Since no experimental constraints have been
obtained for the new couplings,
there exists considerable latitude in the theoretical expectation for
$\hpieff$.

For two decades now, the benchmark theoretical calculation of
$\bra{N\pi}\hweak\ket{N}$ has been
the SU(6)$_w$/quark model approach of Ref. \cite{ddh}, updated in Ref.
\cite{fcdh}. We have argued, however, that the DDH/FCDH calculation
does not manifest the general strictures of chiral invariance obtained
in the present analysis. At the quark level, this chiral structure reflects
the importance of
the \lq\lq disconnected" $q\bar q$ components of the sea. While
relativistic quark models contain
$q\bar q$ sea quark effects in the guise of Z-graphs or
lower-component wavefunctions, the most
common \lq\lq quenched" versions do not include explicit disconnected
pairs\footnote{Some effects of
disconnected $q\bar q$ pairs may, however, hide in the effective parameters
of the quark model, such as
the string tension \cite{Gei90}.}. Given the size of the chiral corrections
associated in part with the
disconnected insertions, it may then be not so surprising to find a
possible discrepancy between the experimental value
for $\hpieff$ and the DDH/FCDH \lq\lq best value".

Applying chiral corrections to the SU(3) analysis of $\Delta S=1$
hyperon decays may help to close the
gap between $\hpieff$ and $h_\pi^\sst{DDH}$. Presumably, similar
corrections should be applied in other
approaches not containing explicit pionic degress of freedom. In the longer
run, one may be able to
use the chiral structure of $\hpieff$ to extrapolate an unquenched
lattice calculation with heavy quarks into the physical regime.

\section*{Acknowledgement}
It is a pleasure to thank J.L. Goity and N. Isgur for useful discussions.
This work was supported in part under U.S. Department of Energy contract
\#DE-AC05-84ER40150, the National Science Foundation, and a National Science
Foundation Young Investigator Award.


\newpage

\appendix

\section{PV Lagrangians}


Here we present the full expressions for some of the PV Lagrangians not
included
in the main body of the paper.
The analogues of Eqs. (\ref{n1}-\ref{n3}) are
\begin{eqnarray}\label{d1}\nonumber
{\cal L}^{\pi\Delta N}_{\Delta I=0} &=&f_1 \epsilon^{abc} \bar N i\gamma_5
[X_L^a A_\mu X_L^b +X_R^a A_\mu X_R^b] T_c^\mu \\
&& +g_1 \bar N [A_\mu, X_-^a]_+ T^\mu_a+g_2 \bar N [A_\mu, X_-^a]_- T^\mu_a
+{\hbox{h.c.}}\\
\nonumber && \\
\label{d2} \nonumber
{\cal L}^{\pi\Delta N}_{\Delta I=1} &=&
f_2 \epsilon^{ab3} \bar N i\gamma_5 [A_\mu, X_+^a]_+ T^\mu_b
+f_3 \epsilon^{ab3}\bar N i\gamma_5[A_\mu, X_+^a]_- T^\mu_b \\ \nonumber
&&+{g_3\over 2}\bar N [(X_L^a A_\mu X_L^3-X_L^3 A_\mu X_L^a)-
(X_R^a A_\mu X_R^3-X_R^3 A_\mu X_R^a)] T^\mu_a\\ \nonumber
&&+{g_4\over 2} \{\bar N [3X_L^3 A^\mu (X_L^1 T^1_\mu +X_L^2 T^2_\mu ) +
3(X_L^1 A^\mu
X_L^3 T^1_\mu
+X^2_L A^\mu X^3_L T^2_\mu) \\
&&-2 (X_L^1 A^\mu X_L^1 +X_L^2 A^\mu X_L^2-2X_L^3 A^\mu X_L^3)T^3_\mu]
-(L\leftrightarrow R) \}
+{\hbox{h.c.}} \\
\nonumber && \\
\label{d3} \nonumber
{\cal L}^{\pi\Delta N}_{\Delta I=2} &=&
f_4 \epsilon^{abd} {\cal I}^{cd}\bar N i\gamma_5  [X_L^a A_\mu X_L^b
+X_R^a A_\mu X_R^b]T^\mu_c \\ \nonumber
&&+f_5 \epsilon^{ab3} \bar N i\gamma_5  [X_L^a A_\mu X_L^3+X_L^3 A_\mu X_L^a
+(L\leftrightarrow R)]T^\mu_b \\
&&+g_5 {\cal I}^{ab}\bar N [A_\mu, X_-^a]_+ T^\mu_b
+g_6 {\cal I}^{ab}\bar N [A_\mu, X_-^a]_- T^\mu_b
+{\hbox{h.c.}}\ \ \ ,
\end{eqnarray}
where the terms containing $f_i$ and $g_i$ start off with one- and two-pion
vertices, respectively. In the heavy baryon expansion, the terms containing
the $f_i$ start to contribute at ${\cal O}(1/\mn)$. The leading order term
vanishes since $P_+\cdot i\gamma_5 \cdot P_+ =0$. Since we work only to
lowest order in the $1/\mn$ expansion, we obtain no contribution from the
terms containing the $f_i$.

For the pv $\pi \Delta \Delta$ effective Lagrangians we have
\begin{equation}\label{ddd1}
{\cal L}^{\pi \Delta}_{\Delta I=0} =j_0 \bar T^i A_\mu \gamma^\mu T_i \; ,
\end{equation}
\begin{eqnarray}\label{ddd2}\nonumber
{\cal L}^{\pi \Delta}_{\Delta I=1} ={j_1\over 2} \bar T^i  \gamma^\mu T_i
Tr (A_\mu X_+^3)
-{k_1\over 2} \bar T^i  \gamma^\mu \gamma_5 T_i  Tr (A_\mu X_-^3)&\\ \nonumber
-{h^1_{\pi \Delta }\over 2\sqrt{2}}f_\pi \bar T^i X_-^3 T_i
-{h^2_{\pi \Delta }\over 2\sqrt{2}}f_\pi \{
3T^3 (X_-^1 T^1 +X_-^2 T^2) +3(\bar T^1 X_-^1 +\bar T^2 X_-^2 ) T^3 &\\
\nonumber
-2(\bar T^1 X_-^3 T^1 +\bar T^2 X_-^3 T^2 -2\bar T^3 X_-^3 T^3) \}
+j_2 \{ 3[(\bar T^3 \gamma^\mu T^1 +\bar T^1 \gamma^\mu T^3) Tr (A_\mu
X_+^1)&\\ \nonumber
+ (\bar T^3 \gamma^\mu T^2 +\bar T^2 \gamma^\mu T^3) Tr (A_\mu X_+^2)]
-2(\bar T^1 \gamma^\mu T^1 +\bar T^2 \gamma^\mu T^2 -2\bar T^3 \gamma^\mu T^3 )
Tr (A_\mu X_+^3) \} &\\ \nonumber
+k_2 \{ 3[(\bar T^3 \gamma^\mu\gamma_5 T^1 +\bar T^1 \gamma^\mu\gamma_5
T^3) Tr (A_\mu X_-^1)
+ (\bar T^3 \gamma^\mu\gamma_5 T^2 +\bar T^2 \gamma^\mu\gamma_5 T^3) Tr (A_\mu
X_-^2)] &\\ \nonumber
 -2(\bar T^1 \gamma^\mu\gamma_5 T^1 +\bar T^2 \gamma^\mu\gamma_5 T^2 -2\bar T^3
\gamma^\mu\gamma_5 T^3 ) Tr (A_\mu X_-^3) \} &\\ \nonumber
+j_3 \{ \bar T^a \gamma^\mu [A_\mu, X_+^a]_+ T^3 +
\bar T^3 \gamma^\mu [A_\mu, X_+^a]_+ T^a \} &\\ \nonumber
+j_4 \{ \bar T^a \gamma^\mu [A_\mu, X_+^a]_- T^3 -
\bar T^3 \gamma^\mu [A_\mu, X_+^a]_- T^a \} &\\ \nonumber
+k_3 \{ \bar T^a \gamma^\mu\gamma_5 [A_\mu, X_-^a]_+ T^3 +
\bar T^3 \gamma^\mu\gamma_5 [A_\mu, X_+^a]_+ T^a \} &\\
+k_4 \{ \bar T^a \gamma^\mu\gamma_5 [A_\mu, X_-^a]_- T^3 -
\bar T^3 \gamma^\mu\gamma_5 [A_\mu, X_+^a]_- T^a \}&\; ,
\end{eqnarray}
\begin{eqnarray}\label{ddd3}\nonumber
{\cal L}^{\pi \Delta}_{\Delta I=2} =
j_5 {\cal I}^{ab} \bar T^a \gamma^\mu A_\mu T^b
+j_6 {\cal I}^{ab} \bar T^i
[X_R^a A_\mu X_R^b +X_L^a A_\mu X_L^b]\gamma^\mu T_i & \\ \nonumber
+k_5 {\cal I}^{ab} \bar T^i
[X_R^a A_\mu X_R^b -X_L^a A_\mu X_L^b]\gamma^\mu\gamma_5 T_i &\\
+k_6 \epsilon^{ab3} [\bar T^3 i\gamma_5 X_+^b T^a
+\bar T^a i\gamma_5 X_+^b T^3 ] &\; ,
\end{eqnarray}
where we have suppressed the Lorentz indices of the $\Delta$ field, i.e.,
$\bar T^\nu \cdots T_\nu$. The vertices with $k_i, h_\Delta $ contain two
pions. All other
vertices contain one pion when expanded to the leading order. At first
sight the
leading order term with $k_6$ in (\ref{ddd3}) has no pions. However such a term
cancels its hermitian conjugate exactly. The constants $h_{\pi\Delta}^i$
are the PV $\pi \Delta\Delta$
Yukawa coupling constants.

In Section 2, the leading terms in the above Lagrangians were expressed in
terms of effective $\pi\pi
N\Delta$ and $\pi\Delta\Delta$ coupling constants. These constants may be
expressed in terms
of the $f_i$, $g_i$, $k_i$, $j_i$ and $h^i_{\pi\Delta}$ as follows:
\begin{eqnarray}\nonumber
h_A^{p \Delta^{++} \pi^- \pi^0}=-2g_1 +2g_2-g_3-3g_4-{2\over 3}g_5+{2\over
3}g_6 & \\ \nonumber
h_A^{p \Delta^{++} \pi^0 \pi^-}=2g_1 +g_3+6g_4+{2\over 3}g_5 & \\ \nonumber
h_A^{p \Delta^{+} \pi^0 \pi^0}=-{\sqrt{6}\over 9}(6g_2+9g_4+2g_6) & \\
\nonumber
h_A^{p \Delta^{+} \pi^+ \pi^-}= -{\sqrt{6}\over 9}(-6g_1-9g_4+4g_5+6g_6)&
\\ \nonumber
h_A^{p \Delta^{+} \pi^- \pi^+}= -{\sqrt{6}\over 9}(6g_1-6g_2-4g_5+4g_6)& \\
\nonumber
h_A^{p \Delta^{0} \pi^+ \pi^0}= -{\sqrt{3}\over
9}(6g_1+6g_2-3g_3+9g_4+2g_5+2g_6)& \\ \nonumber
h_A^{p \Delta^{0} \pi^0 \pi^+}= -{\sqrt{3}\over
9}(-6g_1+12g_2+3g_3+18g_4-2g_5-8g_6)& \\ \nonumber
h_A^{p \Delta^{-} \pi^+ \pi^+}= {\sqrt{2}\over 3}(6g_2-9g_4+2g_6)& \\ \nonumber
h_A^{n \Delta^{++} \pi^- \pi^-}= {\sqrt{2}\over 3}(6g_2-9g_4+2g_6)& \\
\nonumber
h_A^{n \Delta^{+} \pi^- \pi^0}= -{\sqrt{3}\over
9}(6g_1+6g_2+3g_3-9g_4+2g_5+2g_6)& \\ \nonumber
h_A^{n \Delta^{+} \pi^0 \pi^-}= -{\sqrt{3}\over
9}(-6g_1+12g_2-3g_3-18g_4-2g_5-8g_6)& \\ \nonumber
h_A^{n \Delta^{0} \pi^0 \pi^0}= -{\sqrt{6}\over 9}(-6g_2+9g_4-2g_6)& \\
\nonumber
h_A^{n \Delta^{0} \pi^+ \pi^-}= -{\sqrt{6}\over 9}(-6g_1+6g_2+4g_5-4g_6)&
\\ \nonumber
h_A^{n \Delta^{0} \pi^- \pi^+}= -{\sqrt{6}\over 9}(6g_1-9g_4-4g_5-6g_6)& \\
\nonumber
h_A^{n \Delta^{-} \pi^+ \pi^0}= 2g_1-2g_2+g_3+3g_4+{2\over 3}g_5-{2\over 3}
g_6& \\
h_A^{n \Delta^{-} \pi^0 \pi^+}= -2g_1-g_3-6g_4 -{2\over 3} g_5&
\end{eqnarray}

\begin{eqnarray}\nonumber
h_\Delta = h^1_{\pi \Delta}+h^2_{\pi \Delta} &\\
\nonumber \\
\nonumber
h_V^{\Delta^{++}\Delta^+}={1\over \sqrt{6}}(j_0+{4\over
3}j_6)-2\sqrt{6}j_2-{2\sqrt{6}\over 3}
(j_3 +j_4 ) +{j_5\over 3\sqrt{6}} & \\ \nonumber
h_V^{\Delta^{+}\Delta^0}={\sqrt{2}\over 3}(j_0+{4\over
3}j_6)-{2\sqrt{2}\over 9} j_5 & \\
h_V^{\Delta^{0}\Delta^-}={1\over \sqrt{6}}(j_0+{4\over
3}j_6)+2\sqrt{6}j_2+{2\sqrt{6}\over 3}
(j_3 +j_4 ) +{j_5\over 3\sqrt{6}} &
\end{eqnarray}
It is interesting to note there is only one independent PV Yukawa coupling
constant $h_\Delta$
for $\pi \Delta \Delta$ interactions.


\section{Vanishing Loop Contributions}


As noted in Section 3, a large number of graphs which nominally contribute
to $\hpieff$
actually vanish up to ${\cal O}(1/\lamchic)$. Here, we summarize the the
reasons why.

Consider first the corrections due to the PV vector $\pi NN$ vertices. For
FIG. 1 (b) we have
\begin{equation}\label{m_b1}
iM_{(b)}=i{g_A^2\over \sqrt{2}F_\pi^3}\tau^+ (h_v^0+{4\over 3}h_V^2
)(v\cdot q)
\int {d^Dk\over (2\pi)^D} {(S\cdot k)^2\over v\cdot k v\cdot (k+q)
(k^2-m_\pi^2)}
\sim {\cal O} (1/m_N\Lambda_\chi^3) \; ,
\end{equation}
where we have used $v\cdot q\sim {\cal O} (1/m_N)$. Since we are working to
leading order
in the $1/\mn$ expansion, this amplitude does not contribute.
The PV vector interactions also appear in Figs. 1(j1,j2). The corresponding
amplitude is
\begin{equation}\label{m_j}
iM_{(j1)+(j2)}=-i{g_A^2\over \sqrt{2}F_\pi^3}\tau^+ (h_v^0+2h_V^1-{8\over
3}h_V^2 )
\int {d^Dk\over (2\pi)^D} {[(S\cdot k), (S\cdot q)]_+\over v\cdot k v\cdot
(k+q) (k^2-m_\pi^2)}
=0 \;
\end{equation}
This integral vanishes because it is proportional
to $[(S\cdot v), (S\cdot q)]_+$, which vanishes because
$S\cdot v=0$.
All other possible insertions of PV vector $\pi NN$ vertices vanish for
similar reasons
as either (\ref{m_b1}) or (\ref{m_j}). In what follows, we refer only to
insertions involving
the PV $\pi NN$ Yukawa and $\pi\pi NN$ axial couplings.

The propagator corrections in
FIG. 1 (g1)-(h2) vanish after integration since their amplitude of (g1)-g2)
goes as
\begin{equation}
\sim h_\pi\int {d^Dk\over (2\pi)^D} {v\cdot k\over  k^2-m_\pi^2} = 0\ \ \ .
\end{equation}
while the amplitude of (h1)-(h2) goes as
\begin{equation}
\sim h_A^i\int {d^Dk\over (2\pi)^D} {S\cdot k\over  k^2-m_\pi^2} = 0\ \ \ .
\end{equation}

The amplitude of FIG. 1 (i1)-(i4) contains a vanishing integral
\begin{equation}
\sim h_\pi \int {d^Dk\over (2\pi)^D}
{S\cdot k\over v\cdot k (k^2-m_\pi^2)}=0 \ \ \ .
\end{equation}

Figs. 1 (j1)-(j2) do not contribute for the PV Yukawa coupling $h_\pi$
due to charge conservation.
The remaining non-zero diagrams are Figs. 1 (a)-(f2) where the insertions
in loops are of the
Yukawa or axial interations. Figs. (f1), (f2) arises from the
insertion of the counter terms of mass and wave function renormalization.
Fig. 1 (e1)-(e2)  and Fig. 2 (c1)-(c2) contribute to the wave function
renormalization in Eq. (\ref{z-n}).

Due to the heavy baryon projection $P_+ \cdot i\gamma_5\cdot P_+ =0$ the
one pion
PV $\pi N \Delta$ vertex does not contribute in the leading order of heavy
baryon
expansion. Hence,  the chiral loop corrections from FIG. 2 (d1)-(g4) are of
higher order.
Fig. 2 (h1)-(j2) vanishes after integration for reasons similar to
(\ref{m_j}).
The remaining, non-vanishing diagrams are discussed explicitly in Section 3.

As pointed out in Section 2,  both PC and PV two-derivative operators which
conserve CP do not contribute to $h_\pi$ renormalization. For
example, there exists one CP-conserving, PV such operator:
\begin{equation}
\label{eq:two}
{1\over \Lambda_\chi}\bar N \sigma^{\mu\nu} [D_\mu A_\nu -D_\nu A_\mu] N\ \ \ .
\end{equation}
After expansion, the leading term starts with three pions. It contributes
via Figure 1 (a),
at the order of $1/\Lambda_\chi F_\pi^3$. Moreover the loop integration yields
a factor $g_{\mu\nu}$ and leads to zero after contraction with
$\sigma^{\mu\nu}$.

Another possibility comes from insertions of PC two-derivative nucleon
pion operators. There are three PC operators which conserve CP:
\begin{equation}
{1\over \Lambda_\chi}\bar N i\gamma_5 D_\mu A^\mu N\; ,
\end{equation}
\begin{equation}
{1\over \Lambda_\chi}\bar N A^\mu A_\mu N\; ,
\end{equation}
\begin{equation}
{1\over \Lambda_\chi}\bar N \sigma^{\mu\nu} [A_\mu, A_\nu] N\; .
\end{equation}
Note the first two operators are symmetric in the Lorentz indices. Only
the last one arises from the antisymmetric operators listed in Eq.
(\ref{eq:twoderivpc}).
The first one starts off with one pion. The relevant Feynman diagrams are
Figure 1 (c1)-(c2), where
the PV vertex is associated with $h_A^i$. Note these diagrams do not
contribute at leading order
of HBChPT due to the presence of the $i\gamma_5$. The remaining two operators
start off with two pions.  The relevant diagrams are Figure 1 (d1)-(d2).
After integration the contribution of the third operator reads
\begin{equation}
\sim h_\pi \epsilon^{\mu\nu\alpha\beta}v_\alpha S_\beta v^\mu q^\nu
m_\pi^2\ln m_\pi /\Lambda_\chi
F_\pi^2 \; .
\end{equation}
So its contribution is zero. In contrast the second operator yields
\begin{equation}
h_\pi (v\cdot q) m_\pi^2 \ln m_\pi /(\Lambda_\chi F_\pi^2)\; .
\end{equation}
Note $v\cdot q\sim 1/m_N$.
So its contribution is of order $1/(\Lambda_\chi^3 m_N)$. In short, none of
the two-derivative
operators contribute to the renormalization of $h_\pi$ at the
order to which we work.


\section{Bag Model Integrals}


Here, we show that the four-quark bag model integrals relevant to the
calculation of
the DDH/FCDH parameters $\eta$ and $y$ have a Taylor expansion in light
quark mass around
$m_q=0$. We write a bag model quark wavefunction as \cite{dgh86,book}
\begin{equation}
\psi(x) = \left( \begin{array}{l} i u(r)\chi \\ \ell(r)
{\vec\sigma}\cdot{\vec r} \chi
 \end{array} \right) \exp(-iEt)\ \ \ ,
\end{equation}
where $\chi$ denotes a two-component Pauli spinor and where wave function
normalization yields
\begin{equation}
\int\ d^3r ( u(r)^2 + \ell(r)^2) = 1 \ \ \ ,
\end{equation}
where the the radial integration runs from 0 to the bag radius, $R$.
The four quark matrix elements of interest here can depend three different
integrals:
\begin{equation}
\label{eq:bagint}
\int\ d^3r u(r)^4\ , \ \ \ \int\ d^3r \ell(r)^4\ , \ \ \ \int\ d^3r\
u(r)^2\ell(r)^2 \ \ \ .
\end{equation}
The quark radial wave functions are
\begin{eqnarray}
u(r) & = &  N j_0({p_n r\over R}) \\
\ell(r) & = & - N \left({\omega_n -m_q R\over \omega_n+m_q R}\right)^{1/2}
j_1({p_nr\over R}) \ \ \ ,
\end{eqnarray}
where
\begin{equation}
\label{eq:tanp}
\tan p_n =- {p_n\over \omega_n +m_q R -1} \  \ \ \  \ (n=1,2,\cdots)
\end{equation}
\begin{equation}
\label{eq:pndef}
 p_n =\sqrt{\omega_n^2-m_q^2 R^2}
\end{equation}
\begin{equation}
\label{eq:N}
N=\sqrt{p_n^4\over R^3 (2\omega_n^2-2\omega_n+m_q R)\sin^2 p_n}
\end{equation}
\begin{equation}
\label{eq:R}
R^4={N\omega_n -Z_0\over 4\pi B}
\end{equation}
B is the bag constant and $Z_0$ is a phenomenological parameter involved with
the center of mass motion of the bag.

For light quarks and lowest eigenmode
\begin{eqnarray}
\omega_0 \approx & (2.043 + 0.493m_q R) \\
N  \approx & 2.27/\sqrt{4\pi R^3} \ \ \ .
\end{eqnarray}

It is straightforward to show that the bag model integrals in Eq.
(\ref{eq:bagint}) have a Taylor expansion about $m_q=0$.  The argument
proceeds by
noting that the quantities $N$,
$R$, $p_n$, $\omega_n$ and the argument of the spherical Bessel functions
all have Taylor series in $m_q$ about $m_q=0$. The existence of this expansion
can be seen be an explicit, iterative construction. First, expand $\omega_n$
and $R$:
\begin{eqnarray}
\omega_n& = & \sum_{n=0}^\infty \omega_{n,k} m_q^k \\
R & = & \sum_{n=0}^\infty R_k m_q^k\ \ \ .
\end{eqnarray}
Now let $m_q=0$ in Eqs. (\ref{eq:tanp},\ref{eq:pndef}). Doing so eliminates
all dependence on $R$ and determines $\omega_{n,0}$. Next, set $m_q=0$ in
Eqs. (\ref{eq:N}, \ref{eq:R}) with $\omega_n\to\omega_{n,0}$. Doing so
determines $R_0$. Now expand Eqs. (\ref{eq:tanp},\ref{eq:pndef}) to first order
in $m_q$. This step yields $\omega_{n,1}$ in terms of $\omega_{n,0}$ and $R_0$.
Expanding Eqs. (\ref{eq:N}, \ref{eq:R}) to first order in $m_q$ then determines
$R_1$ in terms of $\omega_{n,0}$, $\omega_{n,1}$, and $R_0$ and so forth. Note
that at any step of the recursion, the argument of any transcendental
function is
$\omega_{n,0}$. Hence, at any order, a solution for the $\omega_{n,k}$ and
$R_k$ exists.

The expansion of the bag model integrals continues by computing their
derivatives with respect to $m_q$ and using the expansions of $N$, $R$, etc.
in terms of $m_q$ as constructed above.
Taking $n$ derivatives of one of the integrals in Eq. (\ref{eq:bagint})
yields new intregrals
involving powers of $r/R$ times products of the Bessel functions and their
derivatives. Using
the standard Bessel function recursion relations, the derivatives of the
$j_k$ can always be
expressed in terms of other spherical Bessel functions. Since the $j_k$ and
their derivatives are
finite at the origin, and since the radial bag integration is bounded above
by $R$, the
$n$th derivative of any of the integrals in Eq. (\ref{eq:bagint}) is
finite. Thus, each of the
integrals in Eq. (\ref{eq:bagint}) can be expanded in a Taylor series about
$m_q=0$.

\vfill
\eject

{\bf Figure Captions}

\begin{center}
{\sf Figure 1.} {Meson-nucleon intermediate state contributions to the
PV $\pi NN$ vertex $h_\pi$.  The shaded circle denotes the
PV vertex. The solid and dashed lines correspond to the nucleon and pion
respectively.}
\end{center}

\medskip
\begin{center}
{\sf
Figure 2.} {The chiral corrections from $\Delta$ intermediate state, which
is denoted
by the double line.}
\end{center}

\medskip
\begin{center}
{\sf Figure 3.}{Diagrammatic representation of the SU(6)$_w$ components of
$\bra{B'M}\hweak(\Delta S=0,1)
\ket{B}$. Figs. 3a-c correspond, respectively, to $b_{t,v}$, $c_v$, and
$a_{t,v}$. The wavy line
denotes the action of $\hweak$.}
\end{center}

\medskip
\begin{center}
{\sf Figure 4.} {Chiral corrections to the $B\to B' M$ nonleptonic weak decay.}
\end{center}

\medskip
\begin{center}
{\sf Figure 5.} {Quark line diagrams for the renormalization of $h_\pi$ due
to the axial PV
$\pi\pi NN$ interaction. As in Fig. 3, the wavey line denotes the action of
$\hweak$. Fig. 5a
shows a typical contribution to $h_A^i$. Figs 5b,c denote the corresponding
loop corrections
to $h_\pi$. Fig. 5(b) contains the disconnected $q\bar q$ insertions, while
Fig. 5(c) gives a
Z-graph contribution.}
\end{center}

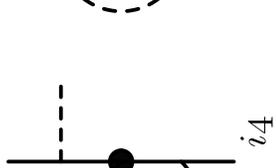
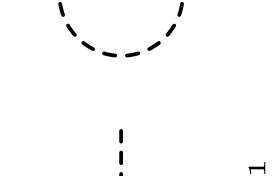
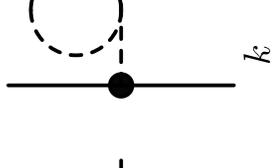
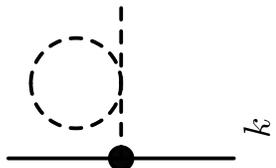
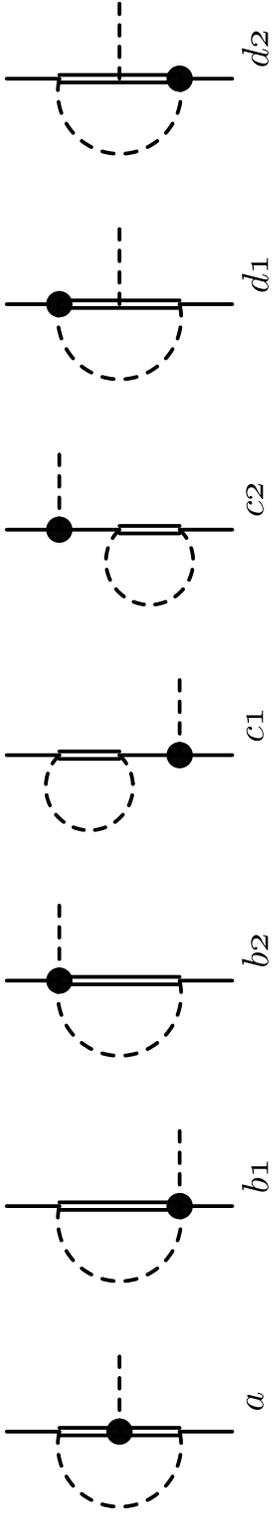
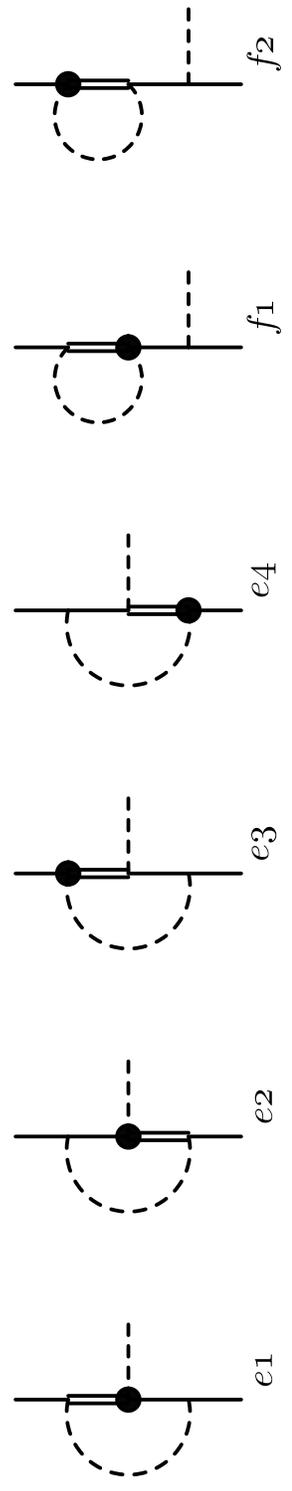
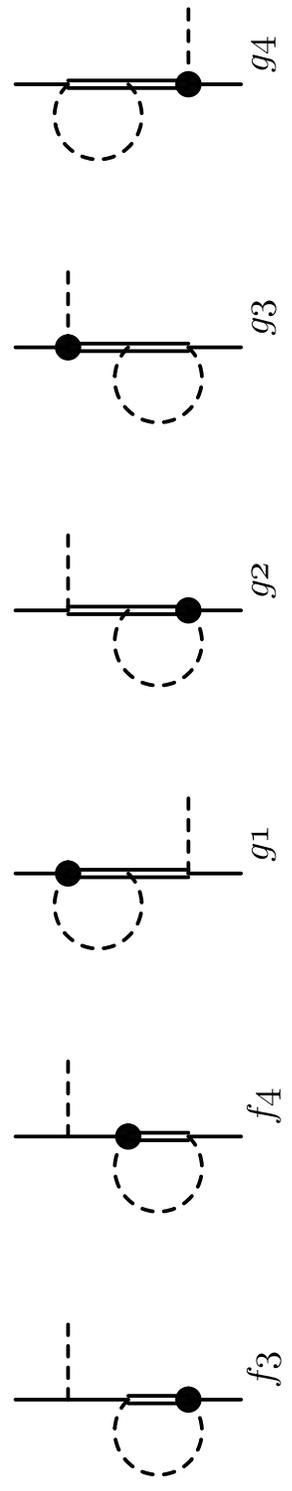
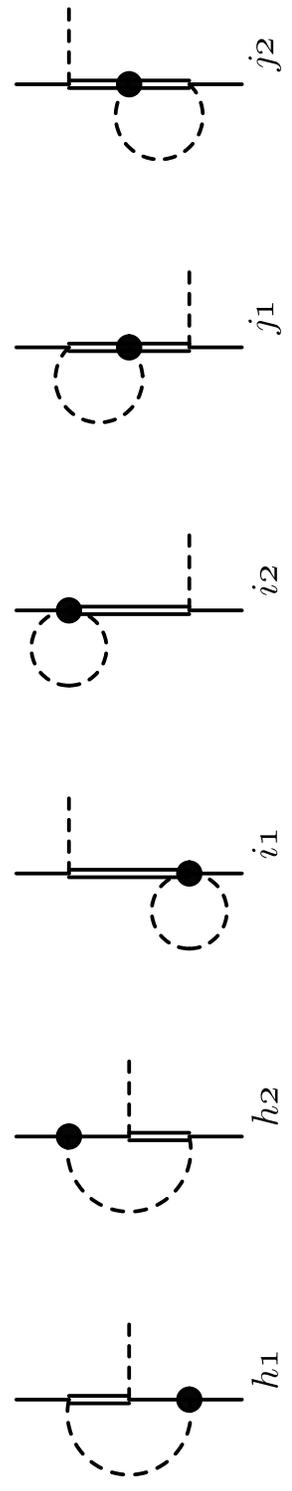
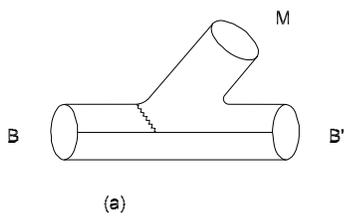
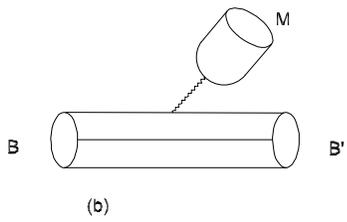
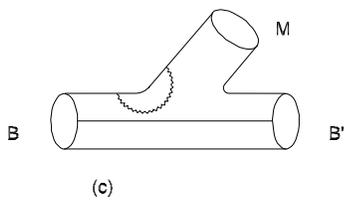
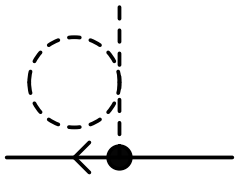
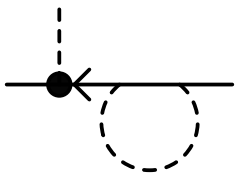
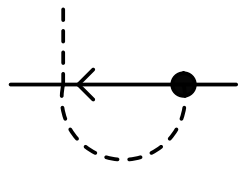
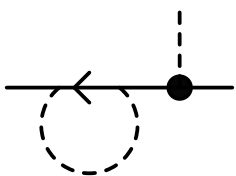
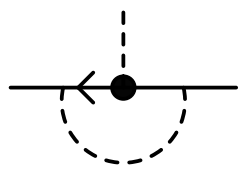
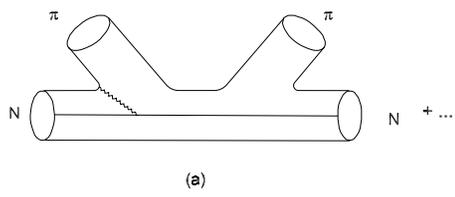
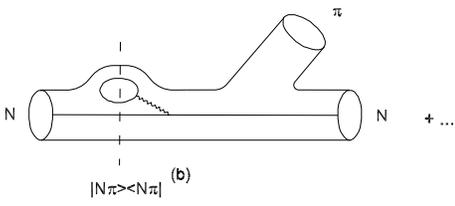
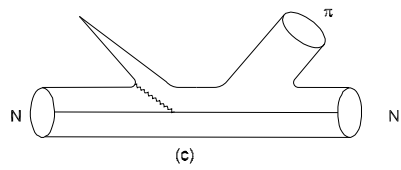
\end{document}